\documentstyle[epsfig]{mn}

\newif\ifAMStwofonts
\ifoldfss
  \newcommand{\rmn}[1] {{\rm #1}}

  \ifCUPmtlplainloaded \else
    \NewTextAlphabet{textbfit} {cmbxti10} {}
    \NewTextAlphabet{textbfss} {cmssbx10} {}
    \NewMathAlphabet{mathbfit} {cmbxti10} {} % for math mode
    \NewMathAlphabet{mathbfss} {cmssbx10} {} %  "   "    "
  \fi
  \ifAMStwofonts
    \ifCUPmtlplainloaded \else
      \NewSymbolFont{upmath} {eurm10}
      \NewSymbolFont{AMSa} {msam10}
      \NewMathSymbol{\upi}     {0}{upmath}{19}
      \NewMathSymbol{\umu}     {0}{upmath}{16}
      \NewMathSymbol{\upartial}{0}{upmath}{40}
      \NewMathSymbol{\leqslant}{3}{AMSa}{36}
      \NewMathSymbol{\geqslant}{3}{AMSa}{3E}

      \let\geq=\geqslant 
    \fi
  \fi
\fi % End of OFSS

\ifnfssone
  \newmathalphabet{\mathit}
  \addtoversion{normal}{\mathit}{cmr}{m}{it}
  \addtoversion{bold}{\mathit}{cmr}{bx}{it}
  \newcommand{\rmn}[1] {\mathrm{#1}}

  \newmathalphabet{\mathbfit} % math mode version of \textbfit{..}
  \addtoversion{normal}{\mathbfit}{cmr}{bx}{it}
  \addtoversion{bold}{\mathbfit}{cmr}{bx}{it}
  \newmathalphabet{\mathbfss} % math mode version of \textbfss{..}
  \addtoversion{normal}{\mathbfss}{cmss}{bx}{n}
  \addtoversion{bold}{\mathbfss}{cmss}{bx}{n}
  \ifAMStwofonts
    \ifCUPmtlplainloaded \else
      \UseAMStwoboldmath
      \makeatletter
      \new@mathgroup\upmath@group
      \define@mathgroup\mv@normal\upmath@group{eur}{m}{n}
      \define@mathgroup\mv@bold\upmath@group{eur}{b}{n}
      \edef\UPM{\hexnumber\upmath@group}
      \new@mathgroup\amsa@group
      \define@mathgroup\mv@normal\amsa@group{msa}{m}{n}
      \define@mathgroup\mv@bold\amsa@group{msa}{m}{n}
      \edef\AMSa{\hexnumber\amsa@group}
      \makeatother
      \mathchardef\upi="0\UPM19
      \mathchardef\umu="0\UPM16
      \mathchardef\upartial="0\UPM40
      \mathchardef\leqslant="3\AMSa36
      \mathchardef\geqslant="3\AMSa3E

      \let\geq=\geqslant 
    \fi
  \fi
\fi % End of NFSS release 1

\ifnfsstwo
  \newcommand{\rmn}[1] {\mathrm{#1}}

  \DeclareMathAlphabet{\mathbfit}{OT1}{cmr}{bx}{it}
  \SetMathAlphabet\mathbfit{bold}{OT1}{cmr}{bx}{it}
  \DeclareMathAlphabet{\mathbfss}{OT1}{cmss}{bx}{n}
  \SetMathAlphabet\mathbfss{bold}{OT1}{cmss}{bx}{n}
  \ifAMStwofonts
    \ifCUPmtlplainloaded \else
      \DeclareSymbolFont{UPM}{U}{eur}{m}{n}
      \SetSymbolFont{UPM}{bold}{U}{eur}{b}{n}
      \DeclareSymbolFont{AMSa}{U}{msa}{m}{n}
      \DeclareMathSymbol{\upi}{0}{UPM}{"19}
      \DeclareMathSymbol{\umu}{0}{UPM}{"16}
      \DeclareMathSymbol{\upartial}{0}{UPM}{"40}
      \DeclareMathSymbol{\leqslant}{3}{AMSa}{"36}
      \DeclareMathSymbol{\geqslant}{3}{AMSa}{"3E}

      \let\geq=\geqslant 
    \fi
  \fi
\fi % End of NFSS release 2

\ifCUPmtlplainloaded \else
  \ifAMStwofonts \else % If no AMS fonts
    \def\upi{\pi}
    \def\umu{\mu}
    \def\upartial{\partial}
  \fi
\fi

\title{Impact of Cosmic Rays on Population III Star Formation}
\author[A. Stacy and V. Bromm]
       {Athena Stacy\thanks{E-mail: minerva@astro.as.utexas.edu} and Volker Bromm \\
 Department of Astronomy, University of Texas, Austin, TX 78712, USA \\}

\pagerange{\pageref{firstpage}--\pageref{lastpage}}
\pubyear{2007}

\begin{document}

\maketitle
\topmargin-1cm

\label{firstpage}

\begin{abstract}
We explore the implications of a possible cosmic ray (CR) background
generated during the first supernova explosions that end the brief lives
of massive Population~III stars. We show that such a CR background could
have significantly influenced the cooling and collapse of primordial gas
clouds in minihaloes around redshifts of $z\sim 15 - 20$, provided the CR
flux was sufficient to yield an ionization rate greater than about
10$^{-19}$ s$^{-1}$ near the center of the minihalo. The presence of  
CRs with energies $\la 10^7$~eV would indirectly enhance the molecular
cooling in these regions, and we estimate that the resulting lower temperatures in these minihaloes would yield a characteristic stellar mass as low as $\sim 10$\,M$_{\odot}$. CRs have a less pronounced effect on the cooling and collapse of primordial gas clouds inside more massive dark matter haloes with virial masses $\ga 10^8$\,M$_{\odot}$ at the later stages of cosmological structure formation around $z\sim10-15$. In these clouds, even without CR flux the molecular abundance is already sufficient to allow cooling to the floor set by the temperature of the cosmic microwave background. 
\end{abstract}


%\indent\indent

\section{Introduction}
How did the cosmic dark ages end, and how was the homogeneous early
Universe transformed into the highly complex state that we observe
today (e.g. Barkana \& Loeb 2001; Miralda-Escud\'{e} 2003; Ciardi \&
Ferrara 2005)?
One of the key questions is to understand the first stars, the so-called Population~III (or Pop~III), since they likely played a crucial role in driving early cosmic evolution  (e.g. Bromm \& Larson 2004; Glover 2005). For instance, the radiation from the first stars contributed to the reionization of the intergalactic medium (IGM) (e.g. Kitayama et al. 2004; Whalen, Abel \& Norman 2004; Alvarez, Bromm \& Shapiro 2006; Johnson, Greif \& Bromm 2007), leading to the end of the cosmic dark ages. After the first stars exploded as supernovae (SNe) they spread heavy elements through the IGM, thereby providing the initial cosmic metal enrichment (e.g. Madau, Ferrara \& Rees 2001; Mori, Ferrara \& Madau 2002; Bromm, Yoshida \& Hernquist 2003; Wada \& Venkatesan 2003; Norman, O'Shea \& Paschos 2004; Greif et al. 2007). Furthermore, this IGM metallicity was likely cucial in governing the transition from early high-mass star formation to the normal-mass star formation seen today (e.g. Omukai 2000; Bromm et al. 2001; Schneider et al. 2002; Bromm \& Loeb 2003; Frebel, Johnson \& Bromm 2007; Smith \& Sigurdsson 2007).

SN explosions are also thought to be the site of cosmic ray (CR) production (e.g. Ginzburg \& Syrovatskii 1969), and the CR background built up from the first SNe may affect the cooling and collapse of primordial gas clouds. Earlier studies (Shchekinov \& Vasiliev 2004; Vasiliev \& Shchekinov 2006) have shown how the presence of CRs at high redshifts could have lowered the minimum mass at which primordial gas could cool and collapse into virialized dark matter (DM) haloes (e.g. Haiman, Thoul \& Loeb 1996; Tegmark et al. 1997). This earlier study assumes the existence of ultra-heavy X particles that decay into ultra-high-energy CRs which interact with the cosmic microwave background (CMB), leading to the production of ionizing photons. In this paper, however, we instead examine CR effects which occur through direct collisional ionization of neutral hydrogen.  The free electrons created from the ionization act as a catalyst for the formation of H$_{2}$ (McDowell 1961). Because molecular hydrogen emits photons through its rovibrational transitions, H$_{2}$ is able to cool primordial gas at temperatures below the threshold for atomic hydrogen cooling ($\la 10^4$~K). Furthermore, CR ionization can indirectly lead to higher abundances of HD, which also acts as a cooling agent in low-temperature primordial gas (see Johnson \& Bromm 2006). However, the presence of CRs can lead to ionization heating, as well. Whether this direct heating effect is strong enough to counter the additional cooling must be determined and will depend on the high-redshift CR energy density. Similar to our work, Rollinde et al. (2005, 2006) used models of early star formation to estimate the CR energy density in the early Universe, constraining it with the observed Li abundances in metal-poor Galactic halo stars. 

In this paper, we will investigate the importance of the CR feedback on Pop III star formation by modeling its effect on the cooling of primordial gas in two cases: collapse within minihaloes, and shocks associated with the virialization of more massive DM haloes during the later stages of structure formation.  The outline for this paper is as follows.  In Section 2 we discuss CR acceleration and propagation in the high-redshift Universe and how these might differ from the present-day case.  Section 3 describes the evolution of primordial gas in minihaloes and in virialization shocks when accounting for the effects of CRs.  For the minihalo case we furthermore discuss how the fragmentation scale could change for a sufficiently high CR flux. We present our conclusions in Section 4.  

\section{Cosmic rays in the high-z Universe}

\subsection{Population III star formation}

Though CR effects will be examined for a range of star formation rates, the typical Pop III star formation rate is taken to be that found in Bromm \& Loeb (2006) at  $z\simeq 15$, which is approximately $\Psi_{*}\simeq 2\times10^{-2}$\,M$_{\odot}$ yr$^{-1}$ Mpc$^{-3}$ in a comoving volume.  This rate was derived using the extended Press-Schechter formalism (Lacey \& Cole 1993) to model the abundance and merger history of cold dark matter (CDM) haloes.  In a neutral medium, before the redshift of reionization, Bromm \& Loeb (2006) assume star formation occurs only in haloes that have become massive enough to enable atomic line cooling with virial termparatures above approximately $10^{4}$~K.  Recent work (see Greif \& Bromm 2006) suggests that these more massive haloes were indeed the dominant site for star formation.  Greif \& Bromm (2006) argue that about 90 percent of the mass involved in metal-free star formation initially cooled through atomic line transitions.  

At higher redshifts such as $z\sim 20$, however, the first stars are thought to have formed inside of $\sim 10^6$~M$_{\odot}$ minihaloes through molecular cooling, and this mode of star formation is much more significant at this time.  For the minihalo case at $z\sim 20$, Yoshida et al. (2003) estimate the rate of star formation through H$_2$ cooling to be $\Psi_{*}\sim 10^{-3}$\,M$_{\odot}$ yr$^{-1}$ Mpc$^{-3}$.  To account for such differences in these determinations of star formation rates, we examine a range of values spanning multiple orders of magnitude.    

%ARS
The Pop III initial mass function (IMF) currently remains highly uncertain, so for this study we do not perform our calculations using a specific IMF.  For simplicity, we instead assume that %ARS 
those Pop~III stars whose masses lie in the pair instability SN (PISN) range (140-260\,M$_{\odot}$) have an average mass of 200\,M$_{\odot}$ (e.g. Heger et al. 2003). A Pop~III initial mass function (IMF) that extends over a large range of masses would imply that only a fraction of these stars were in the PISN range. Thus, here we assume that only slightly less than half of Pop~III stars are in this mass range, leading to a somewhat more conservative value for the CR energy density.  
%ARS
Our estimate generally corresponds to an IMF peaked around 100\,M$_{\odot}$.  %ARS
Due to their high mass, the first stars had very short lifetimes of about 3 Myr (e.g. Bond, Carr \& Arnett 1984), and we therefore assume instantaneous recycling of the stellar material. As described below, this overall picture of Pop~III star formation will be used to estimate the average CR energy density in the high-redshift Universe.

\subsection{Cosmic ray production}

For this paper CRs are assumed to have been generated in the PISNe that may have marked the death of Pop III stars.  The CRs are accelerated in the SN shock wave through the first-order Fermi process by which high-energy particles gain a small percentage increase in energy each time they diffuse back and forth across a shock wave. This yields a differential energy spectrum in terms of CR number density per energy (e.g. Longair 1994):

\begin{equation}
\frac{dn_{\rmn CR}}{d\epsilon} =  \frac{n_{\rmn norm}}{\epsilon_{\rmn min}} \left(\frac{\epsilon}{\epsilon_{\rmn min}}\right)^{x}  \mbox{\ ,}
\end{equation}
   
\noindent where we will use $x=-2$ for definiteness, a typical value given by Fermi acceleration theory (e.g. Bell 1978a). Here, $\epsilon$ is the CR kinetic energy, $\epsilon_{\rmn min}$ is the minimum kinetic energy, and $n_{\rmn CR}$ is the CR number density. Note that our results are somewhat sensitive to the choice of $x$. Choosing values for $x$ that are closer to what is observed in the Milky Way, such as $x=-2.5$ or $x=-3$, or using a CR spectrum similar to that given in Rollinde et al. (2005, 2006) will increase CR heating and ionization, if the overall CR energy density is held constant. For these steeper power laws a higher fraction of the total CR energy resides in the lower-energy CRs, which are the ones that contribute the most to these CR effects, as discussed in Sections 3.1 and 3.4
%ARS
and in the discussion for Figure 3. %ARS
. Such steeper spectral slopes can in fact yield CR heating and ionization rates up to an order of magnitude higher than for $x=-2$.  Using $x=-2$ is therefore a more conservative choice that does not assume any modifications to standard Fermi acceleration theory. 

By equating the total CR energy density $U_{\rmn CR}$ with the integral of the differential CR spectrum over all energies, the normalizing density factor $n_{\rmn norm}$ is estimated to be

\begin{equation}
n_{\rmn norm}=\frac{U_{\rmn CR}}{\epsilon_{\rmn min}{\rmn ln}\left(\frac{\epsilon_{\rmn max}}{\epsilon_{\rmn min}}\right)} 
\approx \frac{1}{10}\frac{\it{U_{\rmn CR}}}{\epsilon_{\rmn min}} \mbox{\ ,}
\end{equation}

\noindent where we get approximately 1/10 for the coefficient choosing $\epsilon_{\rmn max}=10^{15}$~eV and $\epsilon_{\rmn min}=10^{6}$~eV.  The differential energy spectrum is therefore

\begin{equation}
\frac{dn_{\rmn CR}}{d\epsilon} =  \frac{U_{\rmn CR}}{\epsilon_{\rmn min}^{2}\rmn{ln}\left(\frac{\epsilon_{\rmn max}}{\epsilon_{\rmn min}}\right)}\left(\frac{\epsilon}{\epsilon_{\rmn min}}\right)^{-2}   \mbox{\ ,}
\end{equation}

\noindent with

\begin{equation}
U_{\rmn CR}(z)\approx p_{\rmn CR}E_{\rmn SN}f_{\rmn PISN}\Psi_{*}(z)t_{H}(z)(1+z)^3  \mbox{\ .}
\end{equation}

\noindent This can also be written as

\begin{eqnarray}
U_{\rmn CR}(z)\approx2\times10^{-15}{\rmn{\, erg \, cm^{-3}}}
\left(\frac{p_{\rmn CR}}{0.1}\right)\left(\frac{E_{\rmn SN}}{10^{52}\rmn{\, erg}}\right) 
\left(\frac{1+z}{21}\right)^{\frac{3}{2}}\nonumber\\
\times\left(\frac{f_{\rmn PISN}}{2\times10^{-3}\rmn{\, M_{\odot}^{-1}}}\right)
\left(\frac{\Psi_{*}}{2\times10^{-2}\rmn{\,M_{\odot}\, yr^{-1}\, Mpc^{-3}}}\right) \mbox{,}
\end{eqnarray}

\noindent where $p_{\rmn CR}$ is the fraction of SN explosion energy, $E_{\rmn SN}$, that goes into CR energy, and $f_{\rmn PISN}$ is the number of PISNe that occur for every solar mass unit of star-forming material.  Here we take the above star-formation rate to be constant over a Hubble time $t_{H}$, where

\begin{equation}
t_H(z)\simeq 2\times 10^8 {\rmn{\, yr}}\left(\frac{1+z}{21}\right)^{-3/2}  \mbox{\ ,}
\end{equation}

\noindent evaluated at the relevant redshift. We assume that each star quickly dies as a PISN with $E_{\rmn SN}=10^{52}$~erg, appropriate for a 200\,M$_{\odot}$ star 
%ARS
(Heger \& Woosley 2002), %ARS
ten percent of which is transformed into CR energy (e.g. Ruderman 1974). The value of ten percent is derived from Milky Way (MW) energetics, and here we have simply extrapolated this to PISNe.  Very little is known about what value of $p_{\rmn CR}$ applies to PISNe specifically, so assuming their shock structure to be similar to local SNe appears to be a reasonable first guess. We choose 1/500\,M$_{\odot}$ for $f_{\rmn PISN}$, so that there is one PISN for every 500\,M$_{\odot}$ of star-forming material. This implies that somewhat less than half of the star forming mass falls within the PISN range.

Compared to PISNe, the usual core-collapse SNe (CCSNe) thought to accelerate CRs in the Milky Way have an explosion energy that is lower by about an order of magnitude.  However, the masses of their progenitor stars are also much lower, ranging from $\sim 10-40{\rmn \, M}_{\odot}$. Thus, if we have an IMF extending to this lower mass range, there will be approximately an order of magnitude more SNe per unit mass of star-forming material.  However, because $E_{\rmn SN}$ 
%ARS
for low-mass ($\sim 10{\rmn \, M}_{\odot}$) CCSNe progenitors %ARS
will be lower by a similar amount, these two effects may cancel out and result in a total $U_{\rmn CR}$ that is comparable to the PISN case.  Furthermore, the difference in shock velocities, $u_{\rmn sh}$, between a PISN and a CCSN should not be more than a factor of a few. Though CCSNe have a lower explosion energy, this energy is used to accelerate roughly ten times less ejected mass, $M_{\rmn ej}$, than in a PISN. Simple energy conservation, $E_{\rmn SN}\simeq\frac{1}{2}M_{\rmn ej} u_{\rmn sh}^2$, then indicates that the shock velocities of these different SNe will be similar.  This also leads to similar estimates for $\epsilon_{\rmn min}$ for both cases since the minimum CR energy is expected to depend on shock velocity.  For a given star formation rate the results of our study would therefore change little if the source of CRs were CCSNe instead of PISNe.

%ARS
It should be pointed out that the explosion mechanism of the highest-mass CCSNe progenitors, $\sim$ 40\,M$_{\odot}$, is still somewhat uncertain and may be associated with very high explosion energies comparable to that of PISNe, though if most of the CCSNe derive from lower-mass progenitors they should still have lower explosion energies as argued above. Furthermore, we emphasize that the choice of PISNe versus CCSNe is not crucial for our study as long as there exists some source of sufficient CR production. These sources do not necessarily have to be only PISNe. Constraints on the number of PISNe in the early Universe, such as provided by the problem of overproduction of metals in the IGM by PISNe (e.g. Venkatesan \& Truran 2003; Tumlinson et al. 2004; Daigne et al. 2006), will therefore not affect the relevance of this study.  %ARS    

Finally, it is worth noting that the CR energy densities used in this paper are well-within upper limits placed by previous studies. For instance, one of the constraints that can be used to place an upper limit on $U_{\rmn CR}$ is the $^{6}$Li plateau observed in metal-poor Galactic halo stars. While a number ratio of $^{6}$ Li/H$\simeq10^{-11}$ is observed (Asplund et al. 2006), the much smaller ratio $^6$ Li/H$\simeq10^{-14}$ is predicted by big bang nucleosynthesis (BBN). In Rollinde et al. (2005 \& 2006), the overabundance of $^{6}$Li compared to predictions from BBN is assumed to have come from CRs generated by the first stars. In Rollinde et al. (2005) this assumption was used to derive an upper limit on the high-$z$ CR energy density. Assuming a model with a single burst of Pop~III star formation, they found CR energy densities at the time of these bursts to range from approximately $10^{-10}$ to $10^{-12}$ erg cm$^{-3}$ for burst times ranging from $z=100$ to $z=10$. Such limits are roughly three orders of magnitude larger than the $U_{\rmn CR}$ values used in our investigation, so that we do not violate any known constraints on
CR production in the early Universe.

\subsection{High-redshift GZK cutoff}

 The Greisen-Zatsepin-Kuzmin (GZK) cutoff is an upper limit to the energy of a CR if it is extragalactic in origin and thus has traveled through the sea of CMB photons (Greisen 1966; Zatsepin \& Kuzmin 1966). This limit exists because of interactions between CRs and CMB photons which lead to photo-pion and pair production, reducing the energy of the CR for each interaction. The reactions for pair production and photo-pion production, respectively, are (e.g. Berezinsky \& Grigorieva 1988)

\begin{equation}
p^+ + \gamma \rightarrow p^+ + e^{+} + e^{-} \mbox{\ ,}
\end{equation}

\begin{equation}
p^+ + \gamma \rightarrow \pi^+ + n \mbox{\ ,}
\end{equation}

\begin{equation}
p^+ + \gamma \rightarrow \pi^0 + p^+ \mbox{\ .}
\end{equation}

Photo-pion production is the most important interaction for CR protons (De Marco 2005). However, this reaction can only take place if the energy of a CMB photon is above the energy threshold of $\epsilon_{t} =140$~MeV, but this is possible in the rest frame of CRs with sufficiently large Lorentz factors (e.g. Longair 1994).  The reaction will continue until the CR energy falls below the corresponding threshold, which is $\epsilon_{\rmn GZK} = 5\times10^{19}$~eV in today's Universe.  This cutoff has recently been observed by the HiRes experiment (Abbasi et al. 2007). 
   
In the high-redshift Universe, however, the GZK cutoff will be somewhat lower, as can be seen as follows: While in today's Universe the average energy of a CMB photon is $\epsilon_{\rmn CMB} = 2.7 k_{\rmn B} T_{\rmn CMB} = 6\times10^{-4}$ eV, at higher redshifts this energy will be larger by a factor of $(1+z)$. 
In the CR rest frame, the CMB photon energy is

\begin{equation}
\epsilon_{\rmn CMB}' \approx \gamma\mbox{\,} 6\times10^{-4}{\rmn{\, eV}}\,(1+z)  \mbox{\ ,}
\end{equation}

\noindent where $\gamma$ is the Lorentz factor of a CR proton. Equating $\epsilon_{\rmn CMB}'$ with $\epsilon_{t}$, we find

\begin{equation}
\gamma \approx \frac{2\times10^{11}}{(1+z)} \mbox{\ .}
\end{equation}

\noindent We can now calculate the CR energy for which the threshold for photo-pion production is reached:

\begin{equation}
\epsilon_{\rmn GZK}(z) = \gamma m_{\rmn H}c^{2} \approx \frac{3\times10^{20}\rmn{\, eV}}{(1+z)} \mbox{,}
\end{equation}

\noindent where $m_{\rmn H}$ is the mass of a proton. A more precise calculation, carrying out an integration over the entire Planck spectrum and over all angles, yields  

\begin{equation}
\epsilon_{\rmn GZK}(z) = \frac{5\times10^{19}\rmn{\, eV}}{1+z} \approx 2\times10^{18}{\rmn{\, eV}}\left(\frac{1+z}{21}\right)^{-1}  \mbox{.}
\end{equation}

\noindent Thus, at redshifts of 10 or 20, the GZK cutoff is around an order of magnitude smaller than in today's Universe, giving a robust upper limit to the CR energy.  

The GZK cutoff applies only to those CRs that travel large distances through the CMB. A CR can undergo a maximum of approximately 10 interactions with CMB photons before falling below the GZK limit (e.g. Longair 1994). Given an interaction cross section of $\sigma = 2.5 \times 10^{-28}$~cm$^{2}$ for photo-pion production and a CMB density of $n_{\gamma}=5 \times 10^{2}$~cm$^{-3}\, (1+z)^{3}$, the interaction mean-free-path is

\begin{equation}
\lambda = (\sigma n_{\gamma})^{-1} \simeq \frac{10^{25} \rmn{\, cm}}{(1+z)^{3}} \mbox{.}
\end{equation}

\noindent The maximum distance from which ultra-high energy CRs (UHECRs) with  $\epsilon_{\rmn CR} \geq 5\times10^{19}$ eV$/(1+z)$  could have originated is therefore 

\begin{equation}
d_{\rmn max} \simeq 10 \lambda \simeq \frac{10^{26} \rmn{\, cm}}{(1+z)^{3}} \simeq 3 {\rmn{\, kpc}} \left(\frac{1+z}{21}\right)^{-3}\mbox{.}
 \end{equation}

Thus, at $z=20$ a UHECR impinging upon a primordial gas cloud must have been accelerated in a source no more than a proper distance of $\sim 3$~kpc away. The origin of UHECRs remains unknown even in today's Universe, though possible sources range from pulsar winds to active galactic nuclei (AGN) and gamma-ray bursts (GRBs) (e.g. de Gouveia Dal Pino \& Lazarian 2001; Waxman 2001; Torres et al. 2002). Though structures at $z=15$ and $z=20$ are not yet massive enough to form AGN, pulsars and GRBs may be plausible UHECR sources at these redshifts (e.g. Bromm \& Loeb 2006).

\subsection{Magnetic fields}

Magnetic fields are an important component of CR studies due to their effects on both CR acceleration and propagation through the Universe.  For instance, the maximum energy a CR can reach with the Fermi acceleration process has a linear dependence on the ambient magnetic field strength if the growth rate of the CR's energy is limited by its Larmor radius (Lagage \& Cesarsky 1983).  Under this assumption strengths of $\sim 10^{-10}$~G are necessary to accelerate a CR to 10$^{9}$~eV (Zweibel 2003).     

The strength, generation, and dispersal of magnetic fields at high redshifts, however, is still highly uncertain.  One of the most widely held views is that  magnetic seed fields were created soon after the Big Bang and were later amplified in higher-density structures through dynamo mechanisms.  Seed fields are thought to form through various processes including galaxy-scale outflows in the early Universe and the Biermann battery mechanism in regions such as shock waves and ionization fronts (e.g. Kronberg et al. 1999; Gnedin et al. 2000). For one illustrative example, Ichiki et al. (2006) propose a mechanism by which seed fields are created before recombination through second-order cosmological perturbations.  They calculate the seed field at the redshift of recombination, $z_{\rmn rec} \approx 1000$,  to be

\begin{equation}
B_0(z_{\rmn rec})\simeq 10^{-14}\rmn{\, G\,}\left(\frac{\lambda}{10{\rmn \,kpc}}\right)^{-2} \mbox{\ ,}
\end{equation}

\noindent where $\lambda$ is the comoving size of the structure in question for  scales less than about 10\,Mpc. If we take into account magnetic flux freezing as structures become dense and the Universe expands, we get a seed field at virialization of

\begin{equation}
B_{0}(z_{\rmn vir})\simeq 10^{-20}{\rmn{\, G\,}}\left(\frac{\lambda}{10{\rmn \,kpc}}\right)^{-2}\left(\frac{\rho_{\rmn s}}{\rho_{o}}\right)^{2/3}_{z_{\rmn vir}}(1+z_{\rmn vir})^{2} \mbox{\ ,}
\end{equation}

\begin{figure}
\includegraphics[width=.47\textwidth]{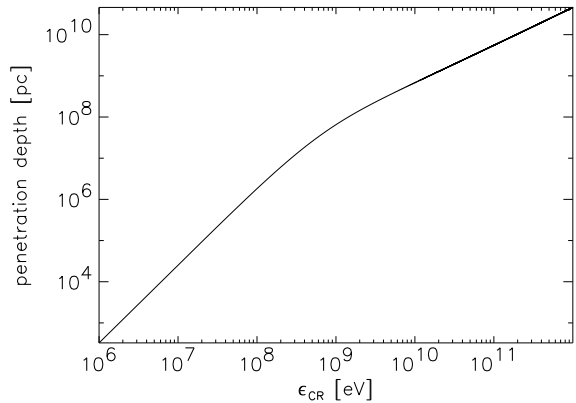}
\caption{Penetration depth as a function of CR energy. We here assume a neutral hydrogen number density of $n_{\rmn H}= 1$~cm$^{-3}$. Only clouds with radii larger than about a few hundred pc will entirely attenuate the lowest-energy CRs through ionization losses. This also implies that low-energy CRs are necessary to contribute significant ionization and heating, as those are the ones that will release most of their energy into gas clouds of size $\sim 0.1 - 1$~kpc.}
\end{figure}

\noindent where $(\rho_{\rmn s}/\rho_{o})_{z_{\rmn vir}}\sim 200$ is the overdensity of the structure in question compared to the average density of the Universe at $z_{\rmn vir}$, the redshift at which the structure first virializes.  For a minihalo with a comoving size of around  10\,kpc and virialization redshift $z_{\rmn vir}=20$, this gives $B_{0}\sim 10^{-16}$\,G. When dynamo effects are taken into account, the magnetic field is further amplified and can grow exponentially on a timescale determined by differential rotation and turbulence within the structure (e.g. Field 1995; Widrow 2002).  Magnetic field amplification within the accelerating region of a SN remnant itself may also increase its strength by up to two orders of magnitude (e.g. van der Lann 1962; Bell \& Lucek 2001).  This magnetic field growth can occur through processes such as flux freezing in the compressed regions of the SN remnant and non-linear amplification through growth and advection of Alfv\'{e}n waves generated by the pressure of CRs themselves.  Within structures that have already experienced star formation, magnetic fields can further be built up through field ejection in stellar winds, SN blastwaves, and protostellar jets (e.g. Machida et al. 2006). 

For these fields to be spread into the general IGM, however, there must be a sufficient degree of turbulent mixing and diffusion in intergalactic regions.  While such processses are effective within structures, it is less obvious that they are also effective in the IGM at eras soon after star formation has begun (Rees 2006).   Thus, the build-up and amplification of magnetic fields in pristine intergalactic matter and in the filaments at high redshifts is very uncertain.  
We can estimate the critical magnetic field at $z\sim 20$ that would influence CR propagation by equating the relevant physical distance scale with the Larmor radius $r_{\rmn L}$ and finding the corresponding magnetic field strength, where for a proton $r_{\rmn L}=\gamma m_{\rmn H} \beta c^2/(e B)$, and $e$ is the proton charge. Considering the maximum scale a CR could travel at $z\sim 20$, we write $r_{\rmn L}=\beta c/H(z=20)$, where $\beta c$ is the CR velocity and $H(z)$ the Hubble constant at redshift $z$.  For $\epsilon_{\rmn CR}=10^6$~eV, we find $r_{\rmn L}\simeq 90$~Mpc, corresponding to a magnetic field strength of 10$^{-20}$~G, several orders of magnitude higher than what Ichiki et al. (2006) estimated for this length scale.  When looking at CR propagation between minihaloes, the physical distance in question is much smaller, $r_{\rmn L}\simeq 1$~kpc.  This corresponds to a magnetic field strength of 5$\times$10$^{-17}$~G, around an order of magnitude stronger than the magnetic field predicted by Ichiki et al. (2006).  These seed fields from recombination are thus too small to affect CR propagation.    

It is therefore not implausible that at high redshift there existed sufficient magnetic fields in the DM haloes to accelerate CRs to relativistic energies while their propagation in the IGM and filaments was very close to rectilinear due to the weak magnetic fields in these regions. This unconfined propagation in the IGM would allow a universal isotropic CR background to build up, and this is what we assume in our analysis.  

\section{Evolution of Primordial Clouds}

\subsection{Idealized models}

\begin{figure*}
\includegraphics[width=.8\textwidth]{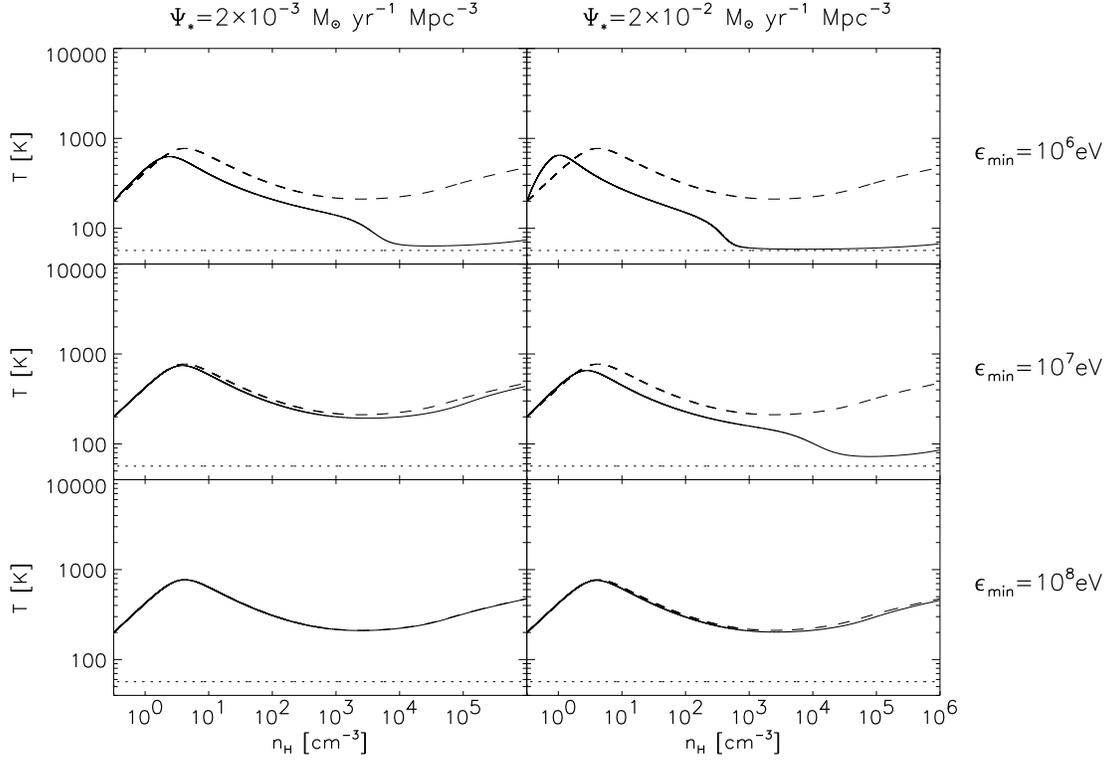}
\caption{Thermal evolution of primordial gas clouds undergoing free-fall collapse inside minihaloes at $z=20$ for various combinations of Pop III star formation rate and $\epsilon_{\rmn min}$. The solid lines show the evolution when the effect of CRs is included, while the dashed lines depict the standard evolution where only cooling due to H$_2$ and HD are considered. The dotted line represents the CMB temperature floor at $z=20$. Notice that the presence of CRs can significantly alter the evolution, in particular lowering the temperature, compared to the canonical picture of how primordial gas behaves in minihaloes. However, this effect crucially relies on the presence of CRs with sufficiently low kinetic energy.}
\end{figure*}

%ARS
\begin{figure*}
\includegraphics[width=.8\textwidth]{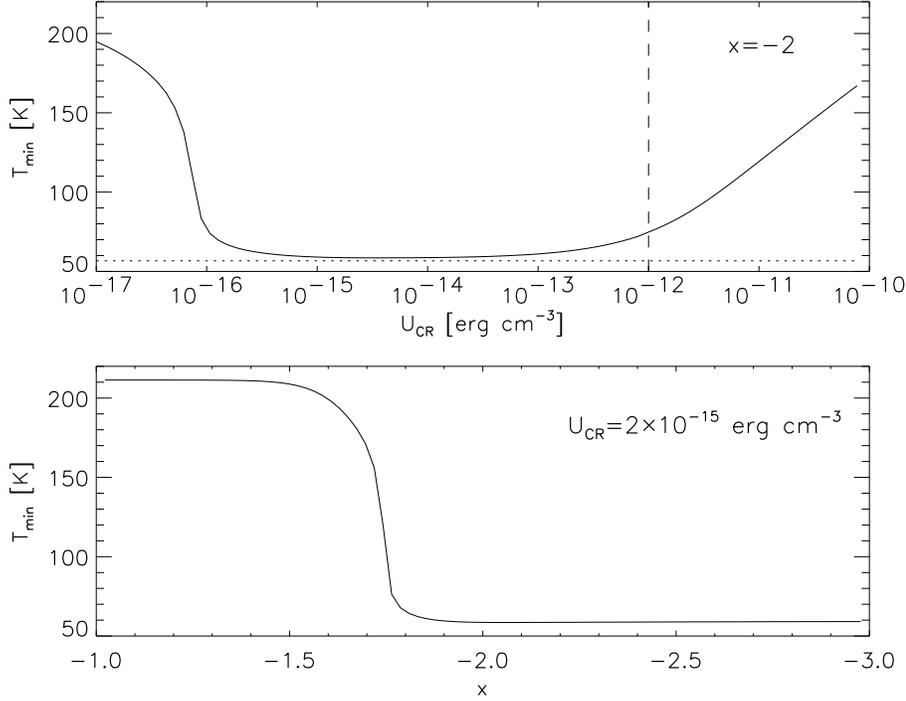}
\caption{Minimum temperature reached by the minihalo gas versus $U_{\rmn CR}$ {\it(top panel)} and spectral slope $x$ {\it (bottom panel)}. The dotted lines denote the CMB temperature, and the dashed line in the upper panel is the upper bound on $U_{\rmn CR}$ based on constraints from the $^6$Li measurements.  Here $\epsilon_{\rmn min}$ is kept constant at 10$^6$ eV.  Note that the cloud cools nearly to the CMB floor for $U_{\rmn CR}$ values ranging over almost four orders of magnitude.  CR-induced heating starts to overcome cooling effects and the minimum temperature begins to rise only once $U_{\rmn CR}$ values are near the $^6$Li constraint.  For shallower spectral slopes, and thus smaller numbers of low-energy CRs for a given $U_{\rmn CR}$, CRs cause little cooling or heating and do not change the minimum temperature of the minihalo gas.}
\end{figure*} %ARS

\begin{figure*}
\includegraphics[width=.8\textwidth]{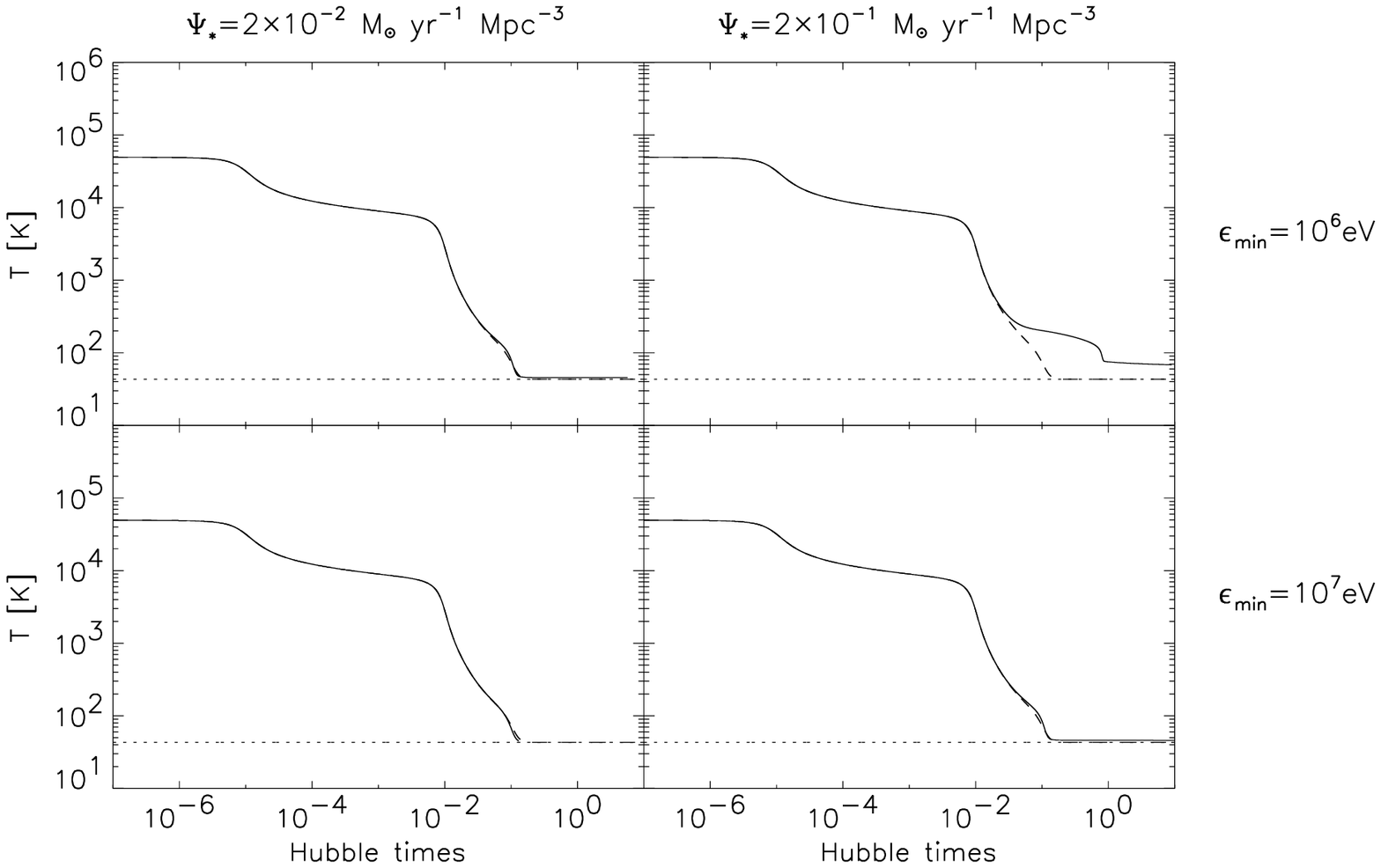}
\caption{Thermal evolution of primordial gas clouds experiencing virialization shocks during the assembly of the first dwarf galaxies at $z=15$. We adopt the same manner of presentation, and the same convention for the lines, as in Figure~2. Notice that the presence of CRs has no impact on the evolution, with the exception of extremely high Pop~III star formation rates {\it (upper right panel)}, where cooling to the CMB floor is prevented.}
\end{figure*}

We examine the evolution of primordial gas by adapting the one-zone models used in Johnson \& Bromm (2006; see also Mackey et al. 2003).  In the $z\sim 20$ collapsing minihalo case, typical halo masses are $\sim 10^{6}$~M$_{\odot}$. The minihaloes form during hierarchical mergers at velocities too low to cause any shock or ionization in the minihalo, so the electron-catalysed molecule formation and cooling is insufficient to allow stars less than around 100\,M$_{\odot}$ to form (e.g. Bromm, Coppi \& Larson 1999, 2002; Abel, Bryan \& Norman 2002). CR ionization and heating, however, can potentially alter the thermal and chemical evolution of primordial gas.  

In our model, the minihalo has an initial ionization fraction of $x_{e}=10^{-4}$ and undergoes free-fall collapse.  Its density therefore evolves according to $dn/dt = n/t_{\rmn{ff}}$, with the free-fall time being

\begin{equation}
t_{\rmn{ff}}=\left(\frac{3\pi}{32G\rho}\right)^{1/2} \mbox{\ ,}
\end{equation}

\noindent where $\rho =\mu m_{\rmn H}n\approx m_{\rmn H}n$ and $\mu =1.2$ is the mean molecular weight for neutral primordial gas.  The initial density was taken to be the density of baryons in DM haloes at the point of virialization (e.g. Clarke \& Bromm 2003)

\begin{equation}
n_{0}\simeq 0.3 {\rmn{\, cm}^{-3}} \left(\frac{1+z}{20}\right)^{3} \mbox{.}
\end{equation}

\noindent The initial temperature of the gas was taken to be 200~K, and the initial abundances were the primordial ones (e.g. Bromm et al. 2002).

As our second case, we consider strong virialization shocks that arise in
later stages of structure formation during the assembly of the first dwarf
galaxies. The corresponding DM haloes virialize at $z\sim 10 - 15$, and have
masses ranging from $\sim10^{8}$ to $\sim10^{10}$~M$_{\odot}$.  
The conditions during the assembly of the first dwarf galaxies show some key differences from the former case. The virial velocities of these DM haloes are much greater, implying merger velocites that are now high enough to create a shock that can partially ionize the primordial gas (Johnson \& Bromm 2006 and references therein). The post-shock evolution is taken to be roughly isobaric (e.g. Shapiro \& Kang 1987; Yamada \& Nishi 1998). We use an initial post-shock temperature of 

\begin{equation}
T_{\rmn ps} = \frac{m_{\rmn H}u_{\rmn sh}^{2}}{3k_{\rmn B}} \mbox{\ .}
\end{equation}    

\noindent For an initial density of $n_{\rmn ps}$, again found from Equation~(19), the temperature and density will follow the relation $T_{\rmn ps}n_{\rmn ps}\simeq T n$.

\subsection{Thermal and chemical evolution}
\begin{figure*}
\includegraphics[width=.8\textwidth]{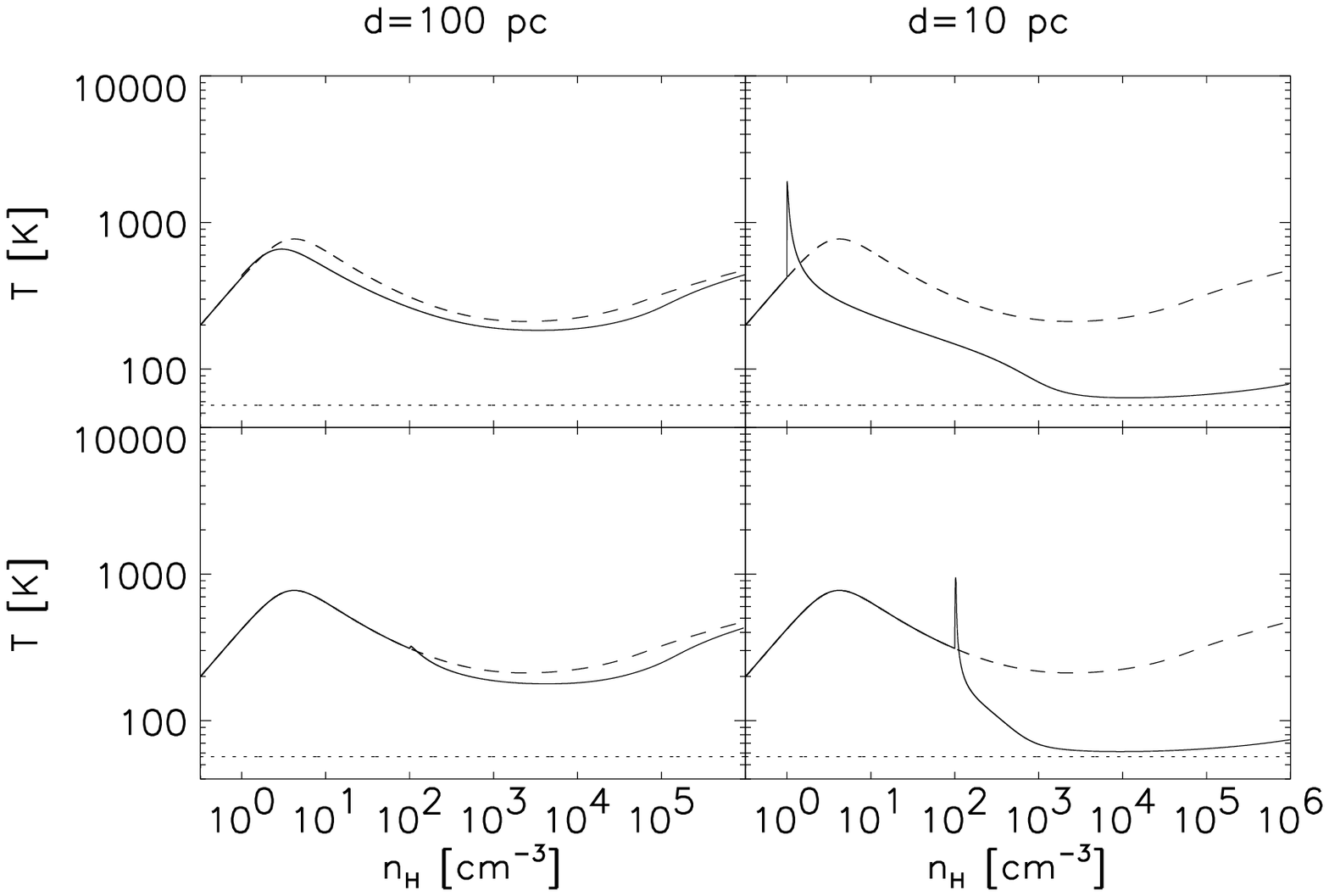}
\caption{Thermal evolution of primordial gas in a minihalo under the effect of a PISN explosion 100~pc away {\it (left column)} and 10~pc away {\it (right column)}. The PISN is assumed to emit CRs for $2\times10^{5}$~yr with a minimum CR energy of 10$^{6}$~eV. We assume that the CR flux reaches the cloud during different stages in its collapse, characterized by the corresponding densities: $n_{\rmn H}=1$~cm$^{-3}$ 
%ARS
{\it (top row)} %ARS
and  $n_{\rmn H}=10^2$~cm$^{-3}$
%ARS
{\it (bottom row)} %ARS
. The dashed lines show the evolution without CR flux prsent. The dotted lines show the CMB temperature floor at $z=20$. Notice that the cloud has to be extremely close to the explosion
site ($d \sim 10$~pc) to experience a significant effect. Such close proximity, however, is very unlikely due to the strong radiative feedback from the PISN progenitor.}
\end{figure*}

In calculating the 
evolution of the primordial clouds, we solve the comprehensive chemical reaction network for all the species included in Johnson \& Bromm (2006), and consider cooling due to H, H$_{2}$, and HD.  The temperature of the CMB sets the lower limit to which the gas can cool radiatively (e.g. Larson 1998).  Assuming that CRs with the above energy spectrum impinge upon the primordial gas cloud, we add the respective heating and ionization rates.  Once a low-energy CR enters the high-density region of the cloud, each time it ionizes an H atom, an electron with average energy $<E>=35$~eV is released (Spitzer \& Tomasko 1968).  Including the ionization energy of 13.6~eV implies that a CR proton loses approximately 50~eV of kinetic energy upon each scattering. This places a limit on the number of scatterings a CR can undergo in a cloud as well as a limit on the distance into the cloud that it can reach before it loses all its energy to ionization.  This distance can be described by a penetration depth 

%\begin{equation}
%D_{p}=N_{sc}^{1/2}\lambda=(\frac{\epsilon_{CR}}{50\rmn{eV}})^{1/2}(n_{H}\sigma(\epsilon))^{-1}
%\end{equation}

\begin{equation}
D_{p}(\epsilon)\approx \frac{\beta c \epsilon}{-\left(\frac{d\epsilon}{dt}\right)_{\rmn ion}}  \mbox{\ ,}
\end{equation}

\noindent where (Schlickeiser 2002) %N$_{sc}$ is the total number of scatterings that a CR of a given energy can undergo, n$_{H}$=1cm$^{-3}$ is the average density of the primordial gas cloud, and $\sigma$ is the ionization cross section for high-energy CRs given by the Bethe-Bloch formula (Bethe 1933), 

%\begin{equation}
%\sigma=\frac{1.23\times10^{-20} Z^{2}}{\beta^{2}}(6.20 + \rmn{log}\frac{\beta^{2}}{1-\beta^{2}} - 0.43\beta^{2})   \rmn{cm}^{2},
%\end{equation}

\begin{equation}
-\left(\frac{d\epsilon}{dt}\right)_{\rmn ion}=1.82\times10^{-7}{\rmn{eV\, s^{-1}}} {n_{\rmn H^0} f(\epsilon)} \mbox{,}
\end{equation}

%\noindent where Z=1 is the charge of the CR, and

\begin{equation}
f(\epsilon)=(1+0.0185\rmn{\,ln}\beta)\frac{2\beta^{2}}{\beta_{0}^{3}+2\beta^{3}} \mbox{,}
\end{equation}

\noindent and

\begin{equation}
\beta=\left[1-\left(\frac{\epsilon}{m_{\rmn H}c^2} +1\right)^{-2}\right]^{1/2}
\mbox{\ .}
\end{equation}

\noindent Here, $m_{\rmn H}$ is the mass of a proton, $\beta=v/c$, and (d$\epsilon$/dt)$_{\rmn ion}$ is the rate of CR energy loss due to ionization. The cutoff value of $\beta_{0}\simeq 0.01$ is appropriate for CRs traveling through a medium of atomic hydrogen, since $\beta_{0}c=0.01 c$ is the approximate orbital velocity of electrons in the ground state of atomic hydrogen. When the velocity of CRs falls below $\beta_{0} c$, the interaction between the CRs and electrons will sharply decrease, as will the ionization rate (Schlickeiser 2002), but for our study all CRs are assumed to be above this critical velocity.  Thus, for a given distance $D$ into a cloud, a CR has an effective optical depth of $D/D_{p}$. Figure~1 shows the dependence of $D_{p}$ on CR energy for a neutral hydrogen density of $n_{\rmn{H^0}}=1$~cm$^{-3}$, which is typical for densities in the minihalo case. Densities will of course greatly increase towards the end of the free-fall evolution, but the size of the collapsing gas cloud, and thus the distance CRs must travel, will decrease. Although the overall attenuation would be slightly greater if the time-dependent density evolution were accounted for instead of using a constant attenuation value, when comparing these two cases the difference is not large enough to yield a significant variation in the minihalo's temperature evolution.  For simplicity, only the typical density was therefore used in calculating the attenuation.

As can be seen in Figure~1, the lowest-energy CRs do not get attenuated until they travel a distance of about a few hundred pc, so in gas clouds of this size or smaller the CR flux will not be significantly attenuated.  This also shows that the low-energy CRs are the ones that will have the greatest ionization and heating contribution to the cloud, as they more readily release their energy into the gas.  In contrast, higher energy CRs will quickly travel through a minihalo without transferring much of their energy into the gas.  They instead lose energy more slowly over much longer distances.  Accounting for the attenuation yields CR ionization and heating rates of

\begin{equation}
\Gamma_{\rmn CR}(D)=\frac{E_{\rmn heat}}{50\rmn \,eV}\int_{\epsilon_{\rmn min}}^{\epsilon_{\rmn max}}
{\left(\frac{d\epsilon}{dt}\right)_{\rmn ion}\frac{dn_{\rmn CR}}{d\epsilon}e^{-D/D_{p}}\rmn{d}\epsilon}
\end{equation}

\noindent and

\begin{equation}
\zeta_{\rmn CR}(D)=\frac{\Gamma_{\rmn CR}}{{n_{\rmn H^0}} E_{\rmn heat}} \mbox{\ .}
\end{equation}
These rates can also be written as

\begin{eqnarray}
\Gamma_{\rmn CR}(D)&=&5\times10^{-29}{\rmn{\, erg \, cm^{-3}\,s^{-1}}}
\left(\frac{U_{\rmn CR}}{2\times10^{-15}{\, \rmn erg \, cm^{-3}}}\right)\nonumber\\
&&\times\left(\frac{E_{\rmn heat}}{6 \rmn \, eV}\right)\left(\frac{n_{\rmn H^0}}{1\rmn{\, cm}^{3}}\right)
\left(\frac{\epsilon_{\rmn min}}{10^{6}\rmn{\, eV}}\right)^{-1}
I(\epsilon)
\end{eqnarray}

\noindent and

\begin{eqnarray}
\zeta_{\rmn CR}(D)&=&5\times10^{-18}{\rmn{\,s}^{-1}}\left(\frac{U_{\rmn CR}}{2\times10^{-15}{\, \rmn erg\, cm^{-3}}}\right)\nonumber\\
&&\times\left(\frac{\epsilon_{\rmn min}}{10^{6}\rmn{\, eV}}\right)^{-1}
I(\epsilon) \mbox{,}
\end{eqnarray}

\noindent where

\begin{equation}
I(\epsilon)=\frac{\int_{\epsilon_{\rmn min}}^{\epsilon_{\rmn max}}f(\epsilon)\left(\frac{\epsilon}{\epsilon_{\rmn min}}\right)^{x}
e^{-D/D_{p}}{\rmn d}\epsilon}
{\int_{\epsilon_{\rmn min}}^{\epsilon_{\rmn max}}
\left(\frac{\epsilon}{\epsilon_{\rmn min}}\right)^{x+1}{\rmn d}\epsilon} \mbox{\ .}
\end{equation}

\begin{figure*}
\includegraphics[width=.8\textwidth]{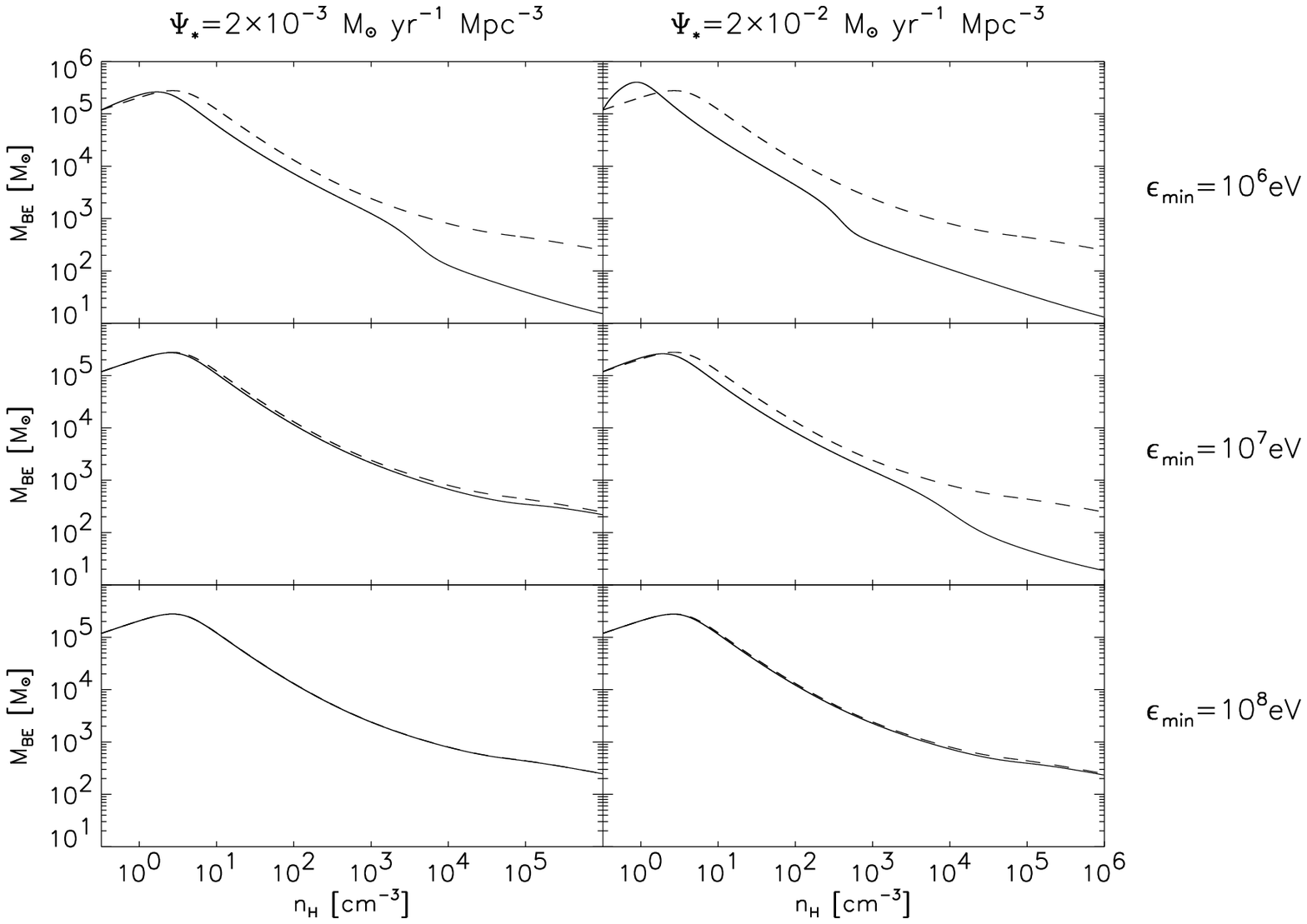}
\caption{$M_{\rmn BE}$ as a function of hydrogen number density for freely-falling clouds inside minihalos at $z\sim 20$. The curves correspond to the evolution shown in Fig.~2, and we adopt the same convention for the lines. Notice that the BE mass, which gives a rough indication for the final stellar mass, can decrease by up to a factor of 10, given that a sufficient flux of low-energy CRs is present.}
\end{figure*}

The factor $E_{\rmn heat}$ is equal to 6~eV for the minihalo case because, though CRs lose about 50~eV of energy after each ionization, only about 6~eV of that energy goes toward heating in a neutral medium (see Spitzer \& Scott 1969; Shull \& van Steenberg 1985).  The value of $E_{\rmn heat}$ increases for media with larger ionization fractions due to an increase in Coulomb interactions between the newly freed electrons and the medium, and for the ionization fractions typical of the virialization shock case, one has $E_{\rmn heat}\simeq 26$~eV.

The minimum kinetic energy, $\epsilon_{\rmn min}$, with which the CRs impinge upon the gas clouds can be roughly estimated by assuming that during Fermi acceleration a particle will acquire the velocity of the shock wave itself after crossing it a single time (see Bell 1978b). This gives a minimum energy $\epsilon_{\rmn min}\sim\frac{1}{2}m_{\rmn H}u_{\rmn sh}^{2}\approx10^{6}$~eV, where $u_{\rmn sh}\approx 10^{4}\rmn{\, km\,s^{-1}}$ is the PISN shock velocity in its initial blast-wave stage.  The effects of changing $\epsilon_{\rmn min}$ will be discussed later.  Due to the power-law distribution of CR energies, the value of the maximum CR kinetic energy, $\epsilon_{\rmn max}$, has less importance since the number densities and ionization rates of highly relativistic CRs are much smaller than those of nonrelativistic and marginally relativistic ones.  We therefore choose a  default value of $\epsilon_{\rmn max}=10^{15}$~eV, a typical maximum value determined from Fermi acceleration theory (e.g. Blandford \& Eichler 1987).

Figure~2 shows the gas cloud evolution for the minihalo collapse case for models with and without CR effects at $z=20$.  The heating rate is evaluated at $D=100$~pc, the characteristic distance to the cloud center, as this is where star formation is expected to take place. Various combinations of $\Psi_{*}=2\times10^{-3}$ or  $2\times10^{-2}\rmn{\, M_{\odot}\, yr^{-1}\, Mpc^{-3}}$ and $\epsilon_{\rmn min}=10^6$, 10$^7$, or 10$^8$~eV are considered.  The resulting CR ionization leads to an increase in the electron abundance in the cloud.  These electrons act as the only catalyst for H$_2$ formation since at this early time in the Universe there are no dust grains on which  H$_2$ could form.  The increased electron abundance thus allows for more  H$_2$ to form.  H$_2$, in turn, can also be used in the main reaction that creates  HD (see, e.g. Johnson \& Bromm 2006).  With the increase in molecular abundance due to CR ionization, the cloud is able to cool to temperatures much closer to the CMB floor if $\Psi_{*}=2\times10^{-2}\rmn{\, M_{\odot}\, yr^{-1}\, Mpc^{-3}}$ and $\epsilon_{\rmn min}=10^6$ or 10$^7$~eV.  This indirect cooling effect is stronger than the direct CR heating effect even for star formation rates up to 100 to 1000 times higher than those shown here.  Futhermore, for $\Psi_{*}$ as low as $2\times10^{-4}\rmn{\, M_{\odot}\, yr^{-1}\, Mpc^{-3}}$, the CR effects are negligible for all $\epsilon_{\rmn min}$ values explored in this study. 

%ARS
To further illustrate how the level of CR-induced cooling can vary with CR parameter values, Figure 3 shows the minimum temperature $T_{\rmn min}$ reached by the minihalo gas versus  $U_{\rmn CR}$ and spectral index $x$ given a constant $\epsilon_{\rmn min} = 10^6$~eV. It is evident that the cloud can cool to temperatures near the CMB floor for $U_{\rmn CR}$ values ranging over nearly four orders of magnitude.  At extremely high values of $U_{\rmn CR}$ that are near the upper limit set by the $^6$Li measurements, CR-induced heating begins to overcome cooling effects, and the minihalo gas can no longer reach temperatures near the CMB floor.  Also, for shallower spectral indices, much less of the CR energy density is distributed at low energies.  For these shallower power laws, CRs thus cause only negligible change in the minimum gas temperature since they yield almost no heating or cooling of the gas.  %ARS

Examination of Figure~2 and the corresponding ionization rates for each case show that CRs can facilitate cooling in the minihalo to nearly the CMB floor if the ionization rate is greater than approximately 10$^{-19}$~s$^{-1}$.  Any rate below this yields negligible cooling. Considering the dependence of the ionization rate on $\epsilon_{\rmn min}$ and $U_{\rmn CR}$ in Equations (28) and (29) for the minihalo case, we can write

\begin{equation} 
\zeta_{\rmn CR} \approx 10^{-19}{\rmn{\, s}^{-1}}\left(\frac{U_{\rmn CR}}{2\times10^{-15}{\rmn \, erg \,cm^{-3}}}\right)\left(\frac{\epsilon_{\rmn min}}{10^{7}\rmn{\,eV}}\right)^{-1.3} \mbox{.}
\end{equation} 

\noindent Recalling the dependence of $U_{\rmn CR}$ on redshift and $\Psi_{*}$ from Equation (4), we find that there will be sufficient ionization, so that the CR-induced cooling will be significant in the minihalo, if the following
condition is met:

\begin{eqnarray}
\left(\frac{\Psi_{*}}{10^{-2}\rmn{\, M_{\odot}\, yr^{-1}\, Mpc^{-3}}}\right)\left(\frac{1+z}{21}\right)^{\frac{3}{2}}
\left(\frac{\epsilon_{\rmn min}}{10^{7}\rmn{\, eV}}\right)^{-1.3}
 \ga 1
\mbox{\ .}
\end{eqnarray}  

\noindent Here, we assume that all other variables that determine $U_{\rmn CR}$ have the same values used thus far in this study.

In Figure~4, we show the evolution of a gas cloud in the virialization shock case at $z=15$.  The heating rate is now evaluated at $D=500$~pc, a typical distance from the outer edge of the halo to its densest region. We again examine the evolution for various combinations of $\Psi_{*}$ and $\epsilon_{\rmn min}$. Note that here we consider star formation rates that are an order of magnitude larger than in the minihalo case. Such increased rates reflect the later formation times of the first dwarf galaxies, so that structure formation has already progressed further. In most of these cases the presence of CRs has a negligible effect. This is in part due to the enhanced CR attenuation resulting from the larger cloud sizes. Furthermore, unlike in the free-fall minihalo evolution, in the case of strong shocks the gas never reaches very high densities of neutral hydrogen. Densities for the shocked case are instead typically around $\la0.2$~cm$^{-3}$ early in the halo's evolution, and 100~cm$^{-3}$ after the gas temperature has reached the CMB floor.  The average heating and ionization rates for this case are thus lower, but the main explanation for the lack of CR-induced differences is that even without CRs the shocked region is highly ionized and able to cool through molecular transitions to the CMB floor.  As shown in Johnson \& Bromm (2006), HD transitions are able to cool the gas down to the CMB temperature within a Hubble time. Thus, with these cooling mechanisms already in place, CRs can only serve to heat the gas, and the CR heating effect can dominate over molecular cooling only for more extreme star formation rates, as can be seen in the upper right panel of Figure~4, where the gas temperature does not reach the CMB floor in
the presence of strong CR heating.

\subsection{Local CR Feedback}

Our study thus far has assumed that the CRs possibly created in the early Universe all become part of a homogeneous and isotropic background.  However, if a particular minihalo is within sufficiently close range to a CR-accelerating PISN, the flux of CRs from the nearby PISN may have a greater effect on the evolution of the minihalo than that from the CR background.

The average CR luminosity associated with the PISN is given by

\begin{eqnarray}
L_{\rmn CR}&=&2\times10^{38}{\, \rmn{erg}}\,{\rmn s^{-1}}\left(\frac{p_{\rmn CR}}{0.1}\right)\nonumber\\
&&\times\left(\frac{E_{\rmn SN}}{10^{52}{\rmn{\,erg}}}\right)
\left(\frac{\Delta t_{\rmn SN}}{2\times10^{5}\rmn{\, yr}}\right)^{-1}  \mbox{\ ,}
\end{eqnarray}

\noindent where $\Delta t_{\rmn SN}$ is the time over which the PISN emits CRs, here assumed to be approximately the time that passes from the beginning of the PISN to the end of its Sedov-Taylor phase of expansion (see Lagage \& Cesarsky 1983). We can then estimate the CR flux, $f_{\rmn CR}$, and energy density, $U_{\rmn CR}$, emitted by the PISN using

\begin{equation}
f_{\rmn CR}=\frac{L_{\rmn CR}}{4\pi d^{2}} \mbox{\ ,}
\end{equation}

\noindent where $d$ is the distance between the CR source and the minihalo.  This can also be expressed as

\begin{eqnarray}
f_{\rmn CR}&\simeq&10^{-4}{\rmn{\, erg}}{\rmn{\, cm^{-2}\, s^{-1}}}\left(\frac{p_{\rmn CR}}{0.1}\right)\nonumber\\
&&\times\left(\frac{E_{\rmn SN}}{10^{52}\rmn{\, erg}}\right)
\left(\frac{\Delta t_{\rmn SN}}{2\times10^{5}\rmn{\, yr}}\right)^{-1}\left(\frac{d}{100\rmn{\,pc}}\right)^{-2} \mbox{.}
\end{eqnarray}

\noindent The CR energy density can now be estimated to be

\begin{equation}
U_{\rmn CR} \simeq \frac{f_{\rmn CR}}{<\beta c>} \mbox{\ .}
\end{equation}

\noindent For $p_{\rmn CR}=0.1$, $E_{\rmn SN}=10^{52}$~erg, and $\Delta t_{\rmn SN}= 2\times$10$^{5}$~yr, $U_{\rmn CR}$ can then be written as

\begin{equation}
U_{\rmn CR} \simeq 5\times10^{-15}{\rmn{erg}}\,{\rmn{cm^{-3}}}
\left(\frac{d}{100\rmn{\,pc}}\right)^{-2}
\frac{\int_{\epsilon_{\rmn min}}^{\epsilon_{\rmn max}}\epsilon^{x}{\rmn d}\epsilon}
{\int_{\epsilon_{\rmn min}}^{\epsilon_{\rmn max}}\beta\epsilon^{x}{\rmn d}\epsilon} \mbox{\ .}
\end{equation}

This energy density is now used in Equations (27) and (28) to determine the CR heating and ionization rates due to the CR emission from a single nearby PISN.  Figure~5 shows the thermal evolution of a minihalo under the influence of the CR flux from a $10^{52}$~erg PISN at 10 and 100~pc away. The PISN was assumed to emit CRs for $2\times10^{5}$~yr with a minimum CR energy of $10^{6}$~eV. The resulting CR flux begins to impinge on the minihalo at times that correspond to two different stages during the collapse, when the gas cloud reaches a density of $n_{\rmn H}=1$~cm$^{-3}$ and when it reaches $n_{\rmn H}=100~$cm$^{-3}$.  At 10~pc, the CR emission allows the gas cloud to cool nearly to the CMB floor, while at 100~pc distance, it causes only a slight change in the thermal evolution of the gas. We also examined the evolution at a distance of 1~kpc from the PISN, but the CR effects were negligible.

For such a local burst of CR emission to have a noticeable impact on the
evolution of the primordial gas, a PISN in very close proximity would be
required. However, such a scenario is not very likely. In effect,
a cloud at 10~pc distance from the explosion would be part of the same
minihalo, and the strong UV radiation from the PISN progenitor would have
evaporated any gas in its vicinity by photoionization-heating (e.g. 
Alvarez et al. 2006; Greif et al. 2007).
When considering realistic cases where the gas evolution in minihaloes could have been significantly impacted by CRs, one therefore needs to invoke a universal background that consists of the global contributions to the CR flux, as opposed to the burst-like emission from a single nearby PISN. However, at close distances the local CR feedback may still provide another source of ionization in nearby gas clouds, indirectly leading to increased molecular cooling, thus helping to facilitate collapse and possibly formation of lower-mass stars, as discussed below in Section~3.5.  

\subsection{Dependence on minimum CR energy}

One of the crucial uncertainties concerning high-redshift CRs is the minimum energy $\epsilon_{\rmn min}$ of the CRs that impinge upon a primordial cloud. Our default value of $\epsilon_{\rmn min}$=10$^{6}$~eV derives from a simple estimate for the lowest possible energy that a CR proton could gain in a SN shock, though other processes may influence this value, possibly increasing or decreasing it. The minimum CR kinetic energy, however, is crucial, because the ionization cross section varies roughly as $\epsilon_{\rmn CR}^{-1}$ for non-relativistic CRs with kinetic energies less than their rest mass energy but greater than $\sim$10$^5$~eV, the energy corresponding to a velocity of $\beta_0\simeq 0.01$.  Thus, higher-energy CRs will travel farther into the cloud before first ionizing a particle.  Only lower-energy CRs will release a large portion of their energy into the cloud, and their absence can significantly lower the overall heating 
%ARS
and ionization %ARS
rate inside the cloud.  This is especially apparent when looking at how the thermal evolution of the minihalo (Figure 2) changes when  $\epsilon_{\rmn min}$ is increased. For typical star formation rates and $\epsilon_{\rmn min}=10^{8}$~eV, the impact of CRs becomes negligible.  
%ARS
It is furthermore interesting to note that for a given $U_{\rmn CR}$ and low $\epsilon_{\rmn min}$ values ($\sim$ 10$^{6}$\,eV), using a powerlaw in momentum instead of a powerlaw in energy will give CR heating and ionization rates about an order of magnitude smaller than those found in the original calculations.  This is because using a momentum powerlaw is equivalent to using a shallower spectral index at low CR energies, resulting in fewer low-energy CRs, and this again illustrates the importance of having a sufficient number of low-energy CRs to contribute to CR effects. %ARS

Decreasing  $\epsilon_{\rmn min}$ to arbitrarily low values, on the other hand, will not yield ever increasing CR ionization rates, as the cross section starts to quickly fall off below $\sim 10^5$~eV,  
%ARS 
and our lowest  $\epsilon_{\rmn min}$ of 10$^6$ eV is already sufficiently close to the peak in ionization cross section to obtain the maximum CR effects.  This was confirmed by actually lowering the value of  $\epsilon_{\rmn min}$ to 1 and 10 keV to see how the heating and ionization rates changed.  In both the minihalo and virialization shock case these rates were changed by less than a factor of two.  %ARS

One of the reasons for the uncertainty in $\epsilon_{\rmn min}$ is that the low-energy part of the CR spectrum is difficult to observe even inside the MW. The interaction of CRs with the solar wind and the resulting deflection by solar magnetic fields prevents any CRs with energies lower than around 10$^8$~eV from reaching Earth, and thus MW CRs below this energy cannot be directly detected.  The Galactic CR spectrum is thought to extend to lower energies between 10$^6$ and 10$^7$~eV (e.g. Webber 1998). However, at non-relativistic energies the CR spectrum in the MW is expected to become much flatter due to ionization energy losses that would be much less relevant in the high-$z$ IGM, as even at the redshifts we consider the density of the IGM is around three orders of magnitude lower than the average MW density.

\subsection{Fragmentation scale}

As is evident in Figure~2, given a strong enough energy density, CRs can serve to lower the minimum temperature that a collapsing cloud inside a minihalo is able to reach. This could have important implications for the fragmentation scale of such gas clouds. We estimate the possible change in fragmentation scale due to CRs by assuming that the immediate progenitor of a protostar will have a mass approximately given by the Bonnor-Ebert (BE) mass (e.g. Johnson \& Bromm 2006)

\begin{equation}
M_{\rmn BE} \simeq 700 {\rmn{\, M_{\odot}}} \left(\frac{T_{\rmn f}}{200{\rmn \,K}}\right)^{3/2} \left(\frac{n_{\rmn f}}{10^{4}{\rmn \,cm^{-3}}}\right)^{-1/2} \mbox{\ ,}
\end{equation}

\noindent where $n_{\rmn f}$ and $T_{\rmn f}$ are the density and temperature of the primordial gas at the point when fragmentation occurs.  For each of the cases shown in Figure 2, the evolution of the BE mass was calculated as the cloud collapsed in free-fall.  This evolution is shown in Figure 6.  For the three cases that had the most significant CR effects, the respective cooling and free-fall times were also examined. When the cooling time, $t_{\rmn cool}$, first becomes shorter than the free-fall time, the gas can begin to cool and fall to the center of the DM halo unimpeded by gas pressure (Rees \& Ostriker 1977; White \& Rees 1978). As the cloud contracts, however, the density increases and the free-fall time becomes shorter, eventually falling back below the cooling time.  At this point, when $t_{\rmn ff}\sim t_{\rmn cool}$, we evaluate the BE mass. This is the instance where the gas undergoes a phase of slow, quasi-hydrostatic contraction, what has been termed `loitering phase' (see Bromm \& Larson 2004). Slow contraction continues, and the gas will leave this loitering regime when its mass is high enough to become gravitationally unstable, thus triggering runaway collapse.  Thus, the `loitering phase' is the characteristic point in the evolution when fragmentation occurs and the BE mass is relevant, at least to zeroth order, given that
star formation is too complex to allow reliable predictions from simple one-zone models such as considered here (e.g. Larson 2003).

Bearing this caveat in mind, we find that, with a sufficiently strong CR flux present, 
the gas density at which the `loitering phase' occurs is increased by a factor of $\sim 10 -100$, and the corresponding temperature decreased by a factor of 2 to 3. Evaluating Equation (37) shows that the BE mass thus decreases by an order of magnitude down to around 10\,M$_{\odot}$. This is the mass scale of what has been termed `Pop II.5' stars (e.g. Mackey et al. 2003; Johnson \& Bromm 2006).  This decrease in fragmentation scale occurs for CR ionization rates greater than $\sim10^{-19}$~s$^{-1}$. A sufficiently large flux of low-energy CRs could thus have facilitated the fragmentation and collapse of primordial minihalo gas into Pop~II.5 stars. Again, we emphasize the somewhat speculative nature of our argument regarding stellar mass scales. However, the general trend suggested here, that the presence of CRs in the early Universe tends to enable lower mass star formation, might well survive closer scrutiny with numerical simulations. 

\section{Summary and Discussion}

We have investigated the effect of CRs on the thermal and chemical evolution of primordial gas clouds in two important sites for Pop~III star formation: minihaloes and higher-mass haloes undergoing strong virialization shocks. We show that the presence of CRs has a negligible impact on the evolution in the latter case, since the primordial gas is able to cool to the CMB floor even without the help of CR ionization, and the direct CR heating is also unimportant unless extremely high star formation rates are assumed. Thus, the thermal and chemical evolution of haloes corresponding to the first dwarf galaxies is rather robust, and the CR emission from the deaths of previously formed Pop~III stars will not initiate any feedback in these star formation sites.  
The impact of CRs on gas in minihaloes, which would typically have formed 
before the more massive dwarf systems, could have been much more
pronounced, given a sufficiently low $\epsilon_{\rmn min}$ and Pop~III star formation rates that are not too low. In each case, the effect of direct CR heating is weaker than the indirect molecular cooling that follows from the increased ionization due to CRs. The additional molecular cooling induced by the CRs allows the gas in a minihalo to cool to lower temperatures. For CR ionization rates above a critical value of $\sim10^{-19}$~s$^{-1}$, we find that the mass scale of metal-free stars might be reduced to $\sim 10$\,M$_{\odot}$, corresponding to what has been termed `Pop~II.5'. 

If CRs did indeed facilitate Pop~II.5 star formation in $z=20$ minihaloes, the subsequent production of CRs would likely have been significantly reduced. Only one Pop~III star is expected to form per minihalo, and it is plausible that with CRs present only one Pop~II.5 star would be able to form in a given minihalo, as feedback effects of this first star may dissipate so much of the minihalo's remaining gas that no other stars can form in the same halo.
A single Pop~II.5 star per minihalo would imply no increase in the global number density of stars relative to the case of one Pop III star per minihalo. If Pop~II.5 stars ever formed they would all die as lower explosion-energy CCSN.  These stars would have lifetimes around 10$^7$~yr, not significantly longer than for the $\sim 100$\,M$_{\odot}$ stars expected to form in minihaloes when CRs are unimportant, so Pop~II.5 stars would still generate CRs almost instantaneously. However, their contribution to the CR energy density would be an order of magnitude less than that from a PISN due to the reduced explosion energy. This reduction in overall CR flux might be sufficient to prevent further CR-induced Pop~II.5 star formation in minihaloes. Stars formed slightly later could then again be classical Pop~III stars which would quickly restore a high level of CR energy density. The overall impact of CRs in the early Universe is thus difficult to determine,
owing to the intricate feedback between star formation and CR production.
We conclude, however, that CRs could significantly influence primordial
star formation, and it will be important to further explore their role with
more sophisticated numerical simulations.

{}

\label{lastpage}

\end{document}